\documentclass[twocolumn]{aastex7}

\usepackage{graphicx}
\usepackage{dcolumn}
\usepackage{bm}
\usepackage{tabularx}
\usepackage{multirow}
\usepackage{capt-of}
\usepackage{color}
\usepackage{url}
\usepackage[utf8]{inputenc}
\usepackage{placeins}

\usepackage{ulem}

\usepackage{graphicx}
\usepackage{booktabs}
\usepackage{adjustbox}
\usepackage{amsmath}

\newcommand{\phTE}{\textsc{IMRPhenomTEHM}\xspace}

\newcommand{\phTHM}{\textsc{IMRPhenomTHM}\xspace}

\definecolor{dodgerblue}{HTML}{1E90FF}
\definecolor{viennared}{HTML}{DA0A14}
\definecolor{ctorange}{HTML}{FF6C0C}
\definecolor{granadagreen}{HTML}{078931}
\definecolor{wales}{HTML}{ff0038}
\definecolor{valenciacfred}{HTML}{ee3524}
\definecolor{barcelonafcgold}{HTML}{edbb00}
\definecolor{jam}{HTML}{A50B5E}
\definecolor{austriawien}{HTML}{441678}

\AtBeginDocument{%
  \hypersetup{
    citecolor=dodgerblue,
    linkcolor=dodgerblue,   
    urlcolor=dodgerblue}}

\newcommand{\UIB}{Departament de F\'isica, Universitat de les Illes Balears, IAC3 -- IEEC, Crta. Valldemossa km 7.5, E-07122 Palma, Spain}

\newcommand{\ICE}
{Institut de Ci\`encies de l'Espai (ICE, CSIC), Campus UAB, Carrer de Can Magrans s/n, 08193 Cerdanyola del Vall\`es, Spain}

\usepackage{xspace}

\begin{document}

\title{
First eccentric inspiral-merger-ringdown analysis of neutron star-black hole mergers
}

\author[orcid=0000-0001-8278-7406]{Maria de Lluc Planas}
\affiliation{\UIB}
\email[show]{m.planas@uib.cat} 

\author[orcid=0000-0002-0445-1971]{Sascha Husa}
\affiliation{\ICE}
\affiliation{\UIB}
\email[no-show]{sascha.husa@csic.es}

\author[orcid=0000-0002-6874-7421]{Antoni Ramos-Buades}
\affiliation{\UIB}
\email[no-show]{antoni.ramos-buades@uib.es} 

\author[orcid=0000-0003-2648-9759]{Jorge Valencia}
\affiliation{\UIB}
\email[no-show]{jorge.valencia@uib.es}



\begin{abstract}

The gravitational wave event GW200105 was the first confident neutron star-black hole (NSBH) merger identified by the LIGO-Virgo-KAGRA collaboration. A recent analysis~\citep{Morras:2025xfu} with an eccentric precessing waveform model that describes the inspiral phase of the $l=2$ and $m=\{0,\pm2\}$ modes has identified this event as the first NSBH merger with strong evidence of orbital eccentricity. In this paper we perform the first analysis of this event with an aligned-spin eccentric waveform model that describes the full inspiral, merger, and ringdown, includes subdominant harmonics, and is partially calibrated to numerical relativity simulations. This analysis confirms the results and finds evidence in favor of eccentricity even with a log-uniform prior in eccentricity. We also analyze the NSBH events GW200115 and GW230529, completing the analysis of all NSBHs with \phTE, and find that these signal are consistent with vanishing eccentricity. Finally, we briefly discuss computational challenges when performing the analysis with time-domain eccentric waveform models.

\end{abstract}



\section{Introduction} 

The LIGO-Virgo-KAGRA collaboration (LVK) has reported the detection of 
77 GW events, found consistent with compact object mergers with a false alarm rate of less than one per year -- see the latest catalog \citep{gwtc3} from the
first three observing runs of the 
current ground based gravitational-wave (GW) detector network \citep{Aasi_2015,Acernese_2015,kagra_2021}, and \citep{LIGOScientific:2024elc} for an event from the fourth run.
Of these, sixty-nine events have been
identified as binary black holes (BBHs), four as
NSBH signals -- the three analyzed here, plus GW191219, limited by the accuracy of current models for such high mass ratios, two as binary neutron stars (BNSs), and two as either a NSBH or BBH~\citep{GW190814,LIGOScientific:2024elc}. GW190814, previously analyzed with \phTE in \citep{Planas:2025jny}, falls within this last category.
Further events have been detected in public data by other authors, see e.g. \citep{Zackay:2019tzo,Venumadhav:2019lyq,Nitz:2020oeq,Nitz:2021uxj,Koloniari:2024kww}.
Understanding the populations of such objects and their formation channels is one of the main goals of GW astronomy, see paper for recent LVK results~\citep{PhysRevX.13.011048}.

In order to compare models of binary formation and stellar evolution with observational GW data, it is necessary to accurately determine the parameters of the sources, in particular the component masses, spin vectors and the orbital eccentricity, which are commonly referred to as intrinsic parameters, and extrinsic parameters such as the luminosity distance. These parameters are measured by comparing the data recorded by the GW detectors with accurate theoretical models of the waveform using the methods of Bayesian inference~\citep{PhysRevD.91.042003, bilby}.

Stellar mass compact binaries formed through isolated binary evolution are expected to circularize~\citep{Stevenson:2017tfq}, i.e.~shed orbital eccentricity before entering the detector band: At early times, 
processes like mass transfer and common-envelope evolution~\citep{Bethe:1998bn,Belczynski:2001uc,Dominik:2012kk} tend to circularize the orbit before the black holes form. Any residual eccentricity is efficiently radiated away through GW emission~\citep{PhysRev.136.B1224, 
PhysRevD.77.081502}, so field-formed compact binaries are expected to enter the LVK band with negligible eccentricity. This has motivated an initial focus of the field of waveform modeling on the quasi-circular (QC) sector, i.e. the case when orbital eccentricity can be neglected. 
Interactions with other objects at sufficiently late times before the coalescence can however result in retaining orbital eccentricity in the sensitive frequency band of the detectors~\citep{10.1111/j.1365-2966.2011.19023.x, Rodriguez:2015oxa,Chattopadhyay:2023pil}. 
Prominent examples of such channels are 
capture or multi-body interactions in dense star clusters~\citep{Miller:2002pg, Samsing:2013kua, Zevin:2018kzq, Stone:2016ryd, Mahapatra:2025agb} or
Zeipel-Kozai-Lidov oscillations in triple systems~\citep{Wen:2002km, Silsbee:2016djf, Kimpson:2016dgk, VanLandingham:2016ccd, Hoang:2017fvh}.
Detecting eccentricity in GW signals is therefore a crucial step toward identifying coalescing binaries that likely did not form through isolated binary evolution, see
e.g.~\citep{Romero-Shaw:2019itr,Romero-Shaw:2021ual,Romero-Shaw:2022xko,Zevin:2021rtf, Zeeshan:2024ovp}.

Waveform models that are routinely used for data analysis by the LVK Collaboration~\citep{gwtc1,gwtc21,gwtc3}, and which describe the entire waveform from the inspiral to the merger and ringdown (IMR), have recently been extended to include orbital eccentricity \citep{Islam:2021mha, Liu_2023,Gamba:2024cvy, Gamboa:2024hli, Planas:2025feq}, and have been applied to measure orbital eccentricity in events identified as BBH signals~\citep{Romero-Shaw:2019itr,Iglesias:2022xfc, Bonino:2022hkj, Ramos-Buades:2023yhy,Gupte:2024jfe, Planas:2025jny}. The results are not entirely conclusive, as many of the signals are short, and in some cases contaminated by noise glitches (deviations from Gaussianity in the noise), although significant evidence points to orbital eccentricity in the case of the event GW200129~\citep{Gupte:2024jfe,Planas:2025jny}. For the two events identified as BNS mergers~\citep{GW170817, GW190425}, analysis with IMR models is computationally expensive, and analyses with perturbative post-Newtonian (PN) waveforms or reweighting techniques have not found evidence for eccentricity but provided upper limits~\citep{Romero-Shaw:2020aaj,Lenon:2020oza}.
For the NSBH event GW200105~\citep{GW200105}, recent analysis~\citep{Morras:2025xfu} based on a PN eccentric precessing inspiral model which includes the $l=2$ and $m=\{0,\pm2\}$ modes~\citep{morras2025} found a value of \mbox{$e_{20} = 0.145^{+0.007}_{-0.097}$} at a reference frequency of \mbox{$20\,\rm{Hz}$}. This result is a milestone in understanding the population of compact binaries, and likely will be studied in great detail by the community, e.g. to further sharpen the result, to deepen the astrophysical interpretation, to understand waveform systematics, and as a benchmark for data analysis algorithms.

Here we present the first full IMR aligned-spin eccentric parameter estimation (PE) analysis of three NSBHs, GW200105, GW200115~\citep{GW200105}, and GW230529~\citep{LIGOScientific:2024elc}, using the waveform model \phTE~\citep{Planas:2025feq}.
It includes subdominant harmonics, and is calibrated to numerical relativity (NR) simulations in the QC sector. Crucially, it remains computationally efficient, enabling PE for low-mass systems using classical sampling techniques at a moderate computational cost.
\begin{figure}
    \centering
    \includegraphics[width=\linewidth]{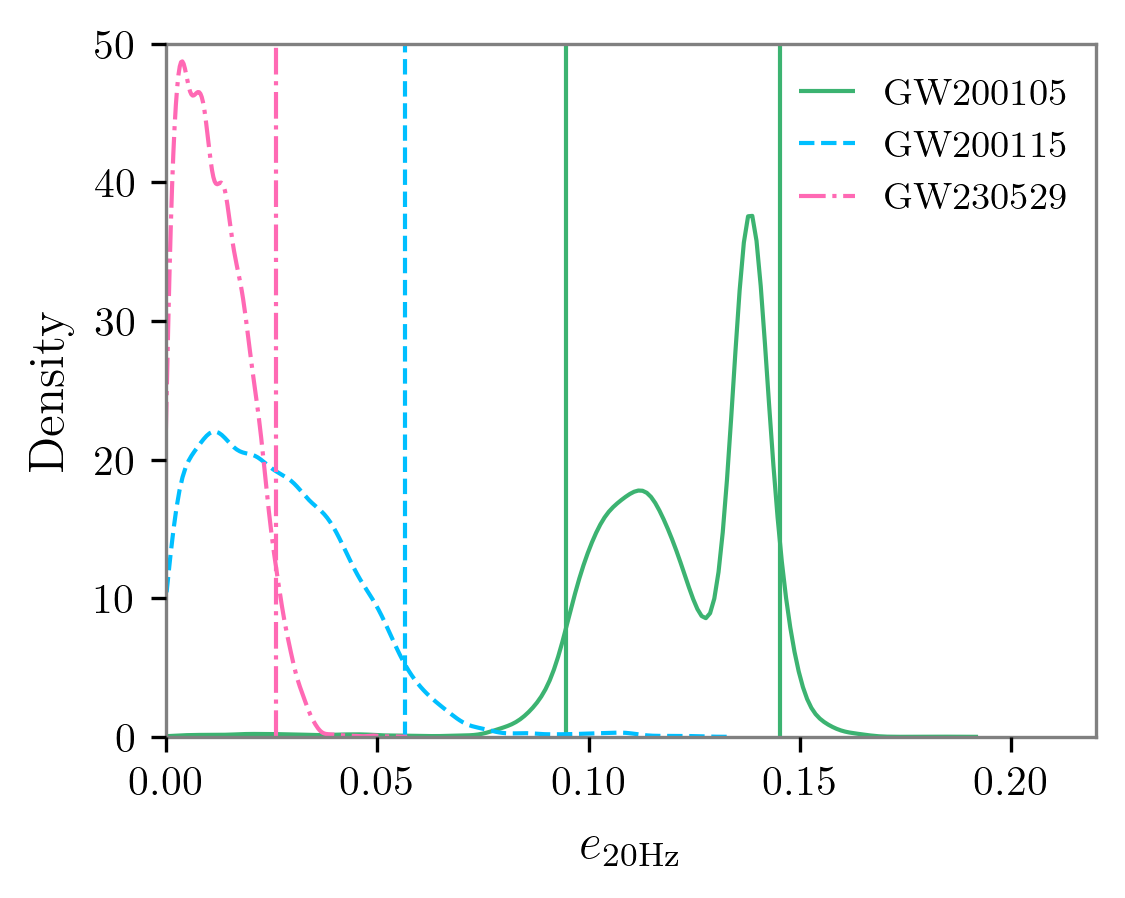}
    \caption{One-dimensional posterior distributions for the orbital eccentricity measured at a reference frequency of \mbox{$20\,\rm{Hz}$} for the NSBH events GW200105 (solid green), GW200115 (dashed blue), and GW230529 (dot-dashed pink) obtained using the \phTE model with a uniform eccentricity prior. Vertical lines indicate the 90\% credible intervals.}
    \label{fig:eccGWs}
\end{figure}
Our main findings are summarized in Fig.~\ref{fig:eccGWs}: Our analysis provides strong evidence for orbital eccentricity in GW200105, while no such support is found for GW200115 or GW230529. 
These results reinforce the finding of eccentricity in \citep{Morras:2025xfu}, now with a fully IMR waveform model with higher multipoles. 
Despite the presence of unexplained features in the eccentricity posteriors for GW200105, we report our results given the novelty and relevance of this analysis. Potential origins of these structures are discussed in the main text and Appendix~\ref{ap:FDecc} discusses eccentric waveform systematics for PE, with a full investigation left for future work.

In Sec.~\ref{sec:methods}, we outline our methods, in Sec.~\ref{sec:results} we describe our results, and we finally address the implications of this work and next steps in Sec.~\ref{sec:conclusions}.

\section{Methodology}\label{sec:methods}

In GW astronomy, the source parameters are determined by Bayesian inference comparing waveform models with the observed data.
The gold standard for waveform accuracy for comparable mass binaries are numerical solutions to the Einstein equations with error estimates. 
Due to their high computational cost, only $\sim10^4$ waveforms exist for the nine-dimensional BBH parameter space~\citep{Boyle:2019kee, Hinder_2019, Hamilton:2023qkv}, with even fewer available for binaries involving neutron stars~\citep{Yamamoto:2008js, Thierfelder:2011yi, PhysRevD.99.103025}.
Waveform models used for PE often incorporate additional theoretical input to improve accuracy, such as PN descriptions of the inspiral~\citep{BlanchetLivRev}, the Effective-One-Body (EOB) formalism~\citep{Buonanno:1998gg, Buonanno:2000ef}, or insights from small mass-ratio perturbation theory~\citep{Barack:2009ux}.

The waveform model we use here to measure eccentricity is \phTE~\citep{Planas:2025feq}, which extends the QC IMR model \phTHM ~\citep{Estelles:2020osj,Estelles:2020twz} to describe GWs from eccentric BBHs.
Besides the dominant $(2,\pm 2)$ spherical harmonic mode, the model also includes $(l,m)= \{(2,\pm 1), (3,\pm 3), (4,\pm 4), (5,\pm 5)\}$. \phTE is not calibrated to eccentric NR waveforms, but through extending \phTHM, it is calibrated to QC NR waveforms up to mass ratio $1/18$.
The model has several limitations: First, as in other current state-of-the-art IMR eccentric models \citep{Liu:2021pkr,Nagar:2021xnh,Gamboa:2024hli,Gamba:2024cvy}, it is assumed that the binary circularizes at the time of merger. 
Additionally, the use of eccentricity-expanded PN approximations for the description of the inspiral limits the applicability of the model to binaries with eccentricities up $e\sim 0.4$ at an orbit-average GW frequency of $10$ Hz, which is safely outside the posterior support for the three events discussed here.
Second, the model does not include matter effects due to the neutron star, in particular tidal effects during the inspiral or disruption effects. For the mass ratios of the events we study here and the values of the primary spin component (see Tab.~\ref{tab:GW_pes} for details), tidal disruption is not expected~\citep{Rantsiou:2007ct,Shibata:2007zm,Etienne:2011ea,Foucart:2010eq,Kyutoku:2011vz,Pannarale:2013jua,Foucart:2018rjc,Foucart:2020ats}. Moreover, at the moderate signal-to-noise ratios (SNR) of these events, tidal effects during the inspiral are not expected to be observable~\citep{Huang:2020pba,GW200105,Gonzalez:2022prs}.
Finally, \phTE assumes spins aligned or anti-aligned with the orbital angular momentum (non-precessing spins). For GW200105 and GW200115, neither the original discovery paper~\citep{GW200105} nor the recent inspiral-only analysis including both eccentricity and spin precession~\citep{Morras:2025xfu} provides evidence for spin precession. 
Similarly, no indications of spin precession were reported for GW230529~\citep{LIGOScientific:2024elc}. Consequently, spin precession is not expected to have a significant impact on our results.

The construction and validation of the model are presented in \citep{Planas:2025feq}. In particular, it was benchmarked against 28 publicly available eccentric NR simulations produced by the Simulating eXtreme Spacetimes (SXS) Collaboration~\citep{Hinder:2017sxy,Boyle:2019kee}, as well as against the spinning eccentric waveform model \textsc{SEOBNRv5EHM}~\citep{Gamboa:2024hli}, which, although more accurate, is computationally restrictive for the analysis considered here. 
In \citep{Planas:2025jny}, we reanalyze 17 BBH GW events using \phTE, finding results consistent with those obtained using \textsc{SEOBNRv4EHM} in combination with the machine learning algorithm \texttt{DINGO}~\citep{dingo} for PE~\citep{Gupte:2024jfe}.
These results indicate that, at current SNRs and for moderate eccentricities, \phTE can be used to reliably measure eccentricity.

We perform Bayesian inference with the \texttt{bilby} framework~\citep{bilby} and the nested sampling algorithm \texttt{dynesty}~\citep{nested_saampling_paper,nested_sampling,dynesty}
to compute posterior distributions for the signal parameters. More specifically, we use \texttt{parallel\_bilby}~\citep{parallel_bilby_paper}, which uses the MPI library~\citep{mpi_mpich} to parallelize across computing nodes to accelerate the runs.
We employ the default nested sampling method, i.e.~using the \texttt{acceptance-walk} method for the Markov Chain Monte Carlo (MCMC) evolution with an average number of accepted steps per MCMC chain of \texttt{naccept=60} and number of live points \texttt{nlive=1000}.

The GW strain data, noise power spectral densities (PSDs) and calibration envelope data are obtained from the public release by the LVK Collaboration~\citep{datarelease, datarelease_230529}, and correspond to those used in the published LVK analyses~\citep{GW200105, LIGOScientific:2024elc}.
For our analysis, we set the starting frequency for likelihood integration to \mbox{$20\,\rm{Hz}$}, consistent with the original studies.
For GW200105, the frequency range \mbox{$46–51\,\rm{Hz}$} is excluded from the Virgo PSD due to additional calibration systematics, following the original LVK studies. LIGO Hanford was not observing at the time of the event and Livingston data underwent a data cleaning procedure to mitigate noise artifacts from scattered light below \mbox{$25\,\rm{Hz}$}.

We sample on a 13-dimensional parameter space representing non-precessing spin BBHs in eccentric orbits. 
The parameters describing the source properties are the components masses $m_i$, with $i=1,2$, the dimensionless spins \mbox{$\chi_i = S_i/m_i^2$}, where $S_i$ are the projections of the component angular momentum vectors along the orbital angular momentum, and finally the
reference orbital eccentricity $e_{20\mathrm{Hz}}$ and mean anomaly $l_{20\mathrm{Hz}}$ at a reference frequency of \mbox{$20\,\rm{Hz}$}. Additionally, we sample on the extrinsic parameters relating the source and detector frames: the coalescence time, coalescence phase, luminosity distance, inclination, polarization angle, right ascension and declination.

The priors on the mass ratio \mbox{$q = m_2/m_1$} with \mbox{$m_1>m_2$} and chirp mass \mbox{$\mathcal{M}= (m_1m_2)^{3/5}/(m_1+m_2)^{1/5}$} are chosen to correspond to a uniform distribution in the component masses. For the spin components $\chi_i$, we use priors corresponding to the projections of a uniform and isotropic spin distribution along a direction perpendicular to the binary's orbital plane~\citep{PhysRevD.91.042003}, with bounds $\chi_i\in[-0.99,0.99]$.
For the luminosity distance $d_L$, we follow the simple prior proportional to $d_L^2$~\citep{gwtc1, gwtc2, gwtc21, gwtc3}, which is justified by the relatively close distance of both events. We set a uniform distribution for the mean anomaly, $l_{\text{20Hz}}\in[0,2\pi]$, while for eccentricity we consider two choices:  \textit{a)} a uniform prior, $e_{\text{20Hz}}\in[0,0.5]$, and \textit{b)} a log-uniform prior with bounds $e_{\text{20Hz}}\in[10^{-4},0.5]$, where the specific lower cutoff of the latter is chosen for consistency with a common choice in the literature. The log-uniform prior expresses ignorance of the order of magnitude of the eccentricity. However, at the present SNR, it would not be possible to measure very small eccentricities, and we consider our uniform prior results as more relevant.

As is standard in GW PE, we perform likelihood integration in the frequency domain starting at \mbox{$20\,\rm{Hz}$}. For time-domain waveforms, this requires starting early enough to ensure all relevant harmonics enter the band by that frequency -- especially important for eccentric signals, which contain multiple mean anomaly harmonics. Appendix~\ref{ap:FDecc} details these considerations and related waveform systematics. By default, we condition the waveform starting three QC cycles before \mbox{$17\,\rm{Hz}$}, and set the reference frequency to \mbox{$20\,\rm{Hz}$}, consistent with~\citep{Morras:2025xfu}. To assess low-frequency effects, we also ran tests for GW200105 with varied starting frequencies and durations.

\section{Results}\label{sec:results}

Our key result is shown in Fig.~\ref{fig:eccGWs}: our analysis for GW200105 finds support of eccentricity \mbox{$e_{20\mathrm{Hz}}=0.12^{+0.02}_{-0.03}$}, similarly as in~\citep{Morras:2025xfu}, with a Bayes factor of \mbox{$\log_{10}\mathcal{B}_{\mathrm{E/QC}}=1.22^{+0.12}_{-0.12}$}, while GW200115 and GW230529 are consistent with QC binaries with eccentricities \mbox{$e_{20\mathrm{Hz}}=0.02^{+0.03}_{-0.02}$} and \mbox{$e_{20\mathrm{Hz}}=0.01^{+0.02}_{-0.01}$}, respectively, supported by Bayes factors \mbox{($\log_{10}\mathcal{B}_{\mathrm{E/QC}}=-0.91^{+0.13}_{-0.13}$)}, and \mbox{($\log_{10}\mathcal{B}_{\mathrm{E/QC}}=-1.36^{+0.12}_{-0.12}$)}.
We verified the robustness of the results for GW200105 presented in Fig.~\ref{fig:eccGWs} by performing PE runs with varying priors and sampler configurations. A selection of these results for each GW event is summarized in Tab.~\ref{tab:GW_pes}, which reports the median values and corresponding 90\% credible intervals for key source properties. For GW200105, the table also includes the official LVK posteriors~\citep{GW200105} obtained using the \textsc{IMRPhenomNSBH} model~\citep{2020PhRvD.101l4059T} for comparison.

\begin{figure*}
    \centering
    \includegraphics[width=\linewidth]{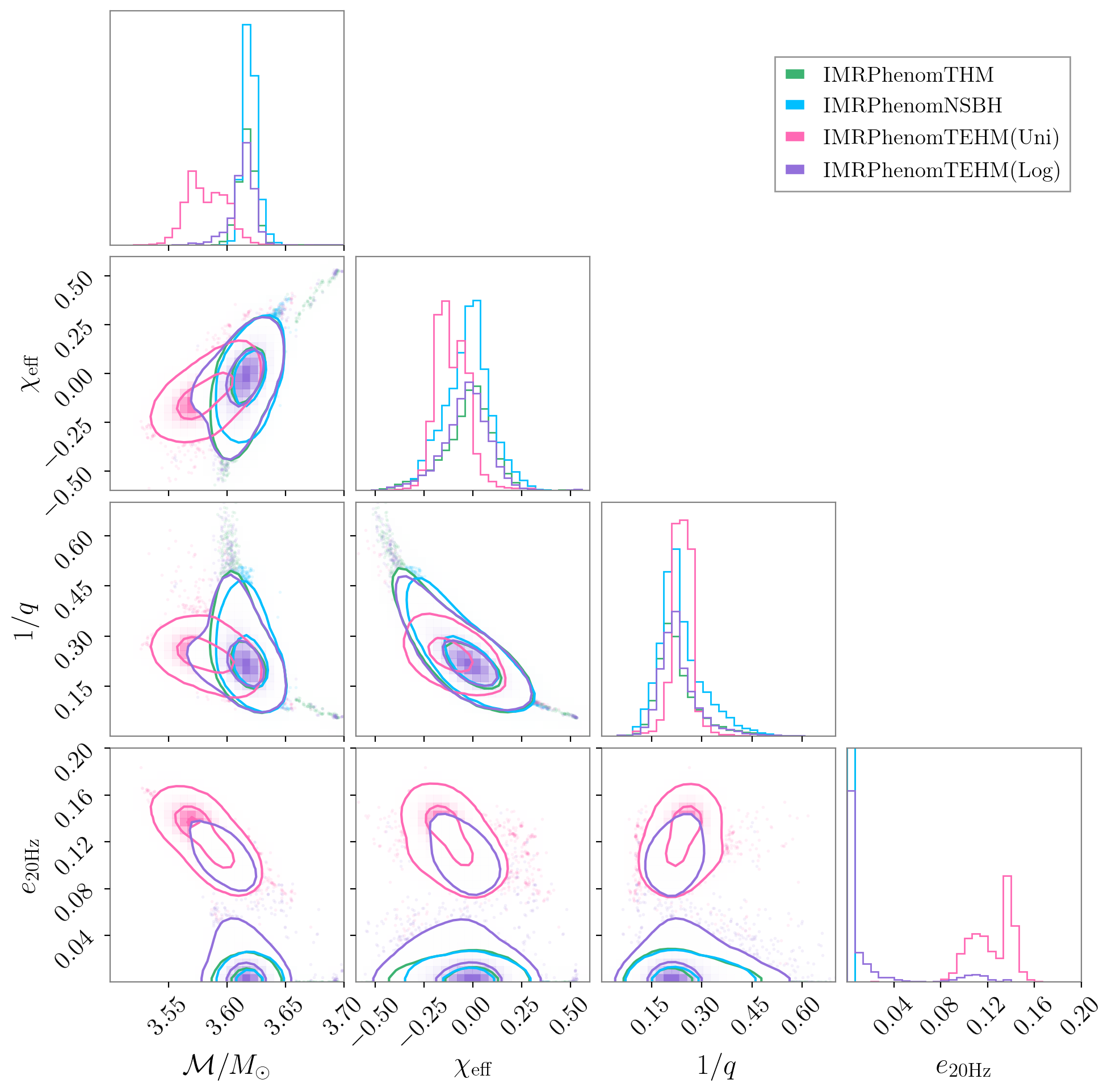}
    \caption{Marginalized one- and two-dimensional posterior distributions for GW200105. We show results from the QC models \phTHM (green) and \textsc{IMRPhenomNSBH} (cyan, taken from the official LVK public release~\citep{GW200105}), as well as the eccentric model \phTE using a uniform prior in eccentricity (pink) and a log-uniform prior (purple). Parameters shown include the chirp mass $\mathcal{M}$, effective spin $\chi_{\mathrm{eff}}$, mass ratio $q$, and eccentricity at \mbox{$20\,\rm{Hz}$} ($e_{20\mathrm{Hz}}$).}
    \label{fig:corner}
\end{figure*}

In Fig.~\ref{fig:corner} we present the GW200105 posterior distributions for some key source properties obtained with two QC models, \phTHM, implemented via the \texttt{phenomxpy} infrastructure~\citep{phenomxpy}, and \textsc{IMRPhenomNSBH}, alongside the eccentric \phTE model using both uniform and log-uniform priors in eccentricity.
We begin by considering the QC waveform models: we find excellent agreement between \phTHM and \textsc{IMRPhenomNSBH} (see also Tab.~\ref{tab:GW_pes} for quantitative results), indicating that the tidal effects do not significantly influence the recovered parameters. While this does not entirely rule out the possibility of degeneracy between tidal and eccentric effects -- both influencing the frequency evolution -- it does suggest that tidal contributions are subdominant and unlikely to induce significant biases in this case.

\begin{deluxetable*}{clccccccccc}[tb]
\tablecaption{Median values and 90\% credible intervals of the posterior distributions for GW200105, GW200115, and GW230529, indicated in each row. The \textsc{IMRPhenomNSBH}~\citep{2020PhRvD.101l4059T} posteriors for GW200105 are taken from the official LVK public release~\citep{GW200105}. The parameters displayed are the total mass $M$ and chirp mass $\mathcal{M}$ (both in solar masses), the mass ratio $q$, the effective-spin parameter $\chi_{\mathrm{eff}}$, the reference eccentricity $e_{20\mathrm{Hz}}$ and mean anomaly $l_{20\mathrm{Hz}}$, the luminosity distance $d_L$, and the network matched-filtered SNR $\mathrm{SNR}^{\mathrm{N}}$. The last column shows the log-10 Bayes factor between the eccentric (E) and the QC aligned-spin (QC) hypothesis $\log_{10}\mathcal{B}_{\mathrm{E}/\mathrm{QC}}$. The spins and eccentric parameters are given at the reference frequency of 20 Hz. \label{tab:GW_pes}}
\tabletypesize{\scriptsize}
\tablehead{
\colhead{\textbf{Event}} & \colhead{\textbf{Model}} & \colhead{$M/M_{\odot}$} & \colhead{$\mathcal{M}/M_{\odot}$} & \colhead{$1/q$} & \colhead{$\chi_\text{eff}$} & \colhead{$e_{20\mathrm{Hz}}$} & \colhead{$l_{20\mathrm{Hz}}$} & \colhead{$d_L$} & \colhead{$\text{SNR}^{\text{N}}$} & \colhead{$\log_{10}\mathcal{B}_{\mathrm{E}/\mathrm{QC}}$}
}
\startdata
\multirow{8}{*}{\texttt{GW200105}} 
& THM            & $11.49^{+1.96}_{-2.04}$ & $3.62^{+0.01}_{-0.01}$ & $0.22^{+0.17}_{-0.07}$ & $-0.01^{+0.17}_{-0.27}$ & -- & -- & $275^{+108}_{-112}$ & $13.81^{+0.18}_{-0.31}$ & -- \\
& TEHM (Uni)     & $10.83^{+1.06}_{-0.60}$ & $3.58^{+0.03}_{-0.02}$ & $0.24^{+0.04}_{-0.05}$ & $-0.12^{+0.15}_{-0.11}$ & $0.12^{+0.02}_{-0.03}$ & $3.25^{+2.73}_{-2.94}$ & $255^{+104}_{-102}$ & $14.39^{+0.32}_{-0.42}$ & $1.22^{+0.12}_{-0.12}$ \\
& TEHM (LogUni)  & $11.38^{+1.87}_{-1.82}$ & $3.62^{+0.01}_{-0.02}$ & $0.22^{+0.15}_{-0.07}$ & $-0.03^{+0.17}_{-0.24}$ & $0.00^{+0.11}_{-0.00}$ & $3.16^{+2.82}_{-2.86}$ & $276^{+106}_{-112}$ & $13.84^{+0.47}_{-0.31}$ & $0.11^{+0.11}_{-0.11}$ \\
& THM (Uni, 64 s)     & $11.55^{+2.04}_{-1.92}$ & $3.62^{+0.01}_{-0.01}$ & $0.21^{+0.15}_{-0.07}$ & $-0.00^{+0.17}_{-0.24}$ & -- & -- & $272^{+107}_{-112}$ & $13.78^{+0.19}_{-0.32}$  & -- \\
& TEHM (Uni, 64 s)     & $10.73^{+1.40}_{-0.57}$ & $3.58^{+0.03}_{-0.02}$ & $0.25^{+0.04}_{-0.06}$ & $-0.14^{+0.18}_{-0.11}$ & $0.13^{+0.02}_{-0.04}$ & $3.99^{+2.00}_{-3.66}$ & $251^{+107}_{-104}$ & $14.31^{+0.30}_{-0.43}$  & $1.00^{+0.12}_{-0.12}$\\
& TEHM (Uni, 13.3 Hz)     &   $10.76^{+1.55}_{-0.58}$ & $3.58^{+0.03}_{-0.02}$ & $0.25^{+0.04}_{-0.07}$ & $-0.13^{+0.19}_{-0.10}$ & $0.13^{+0.02}_{-0.03}$ & $3.16^{+2.83}_{-2.87}$ & $254^{+107}_{-105}$ & $14.36^{+0.31}_{-0.43}$ & -- \\
& TEHM (Uni, L(L1)=25 Hz)     &  $10.89^{+0.83}_{-0.52}$ & $3.58^{+0.02}_{-0.02}$ & $0.24^{+0.03}_{-0.04}$ & $-0.11^{+0.12}_{-0.10}$ & $0.12^{+0.02}_{-0.03}$ & $3.17^{+2.78}_{-2.86}$ & $258^{+104}_{-105}$ & $14.41^{+0.27}_{-0.39}$   &  -- \\
& \textsc{IMRPhenomNSBH} & $11.34^{+2.28}_{-1.81}$ & $3.62^{+0.01}_{-0.01}$ & $0.22^{+0.15}_{-0.08}$ & $-0.02^{+0.19}_{-0.21}$ & -- & -- & $285^{+116}_{-124}$ & $13.46^{+0.17}_{-0.28}$ & -- \\
\hline
\multirow{2}{*}{\texttt{GW200115}} 
& THM            & $8.06^{+1.61}_{-1.45}$ & $2.58^{+0.01}_{-0.01}$ & $0.22^{+0.19}_{-0.08}$ & $-0.10^{+0.20}_{-0.27}$ & -- & -- & $288^{+139}_{-109}$ & $11.40^{+0.22}_{-0.46}$ & -- \\
& TEHM (Uni)     & $7.72^{+1.79}_{-1.36}$ & $2.58^{+0.01}_{-0.01}$ & $0.25^{+0.24}_{-0.10}$ & $-0.17^{+0.25}_{-0.29}$ & $0.02^{+0.03}_{-0.02}$ & $3.17^{+2.79}_{-2.84}$ & $285^{+142}_{-110}$ & $11.40^{+0.24}_{-0.48}$           & $-0.91^{+0.13}_{-0.13}$ \\
\hline
\multirow{2}{*}{\texttt{GW230529}} 
& THM            & $5.22^{+0.93}_{-0.50}$ & $2.03^{+0.00}_{-0.00}$ & $0.41^{+0.33}_{-0.17}$ & $-0.10^{+0.19}_{-0.14}$ & -- & -- & $212^{+109}_{-104}$ & $11.43^{+0.19}_{-0.29}$ & -- \\
& TEHM (Uni)  & $5.14^{+0.96}_{-0.45}$ & $2.03^{+0.00}_{-0.00}$ & $0.44^{+0.34}_{-0.19}$ & $-0.13^{+0.20}_{-0.14}$ & $0.01^{+0.02}_{-0.01}$ & $3.05^{+2.90}_{-2.75}$ & $210^{+116}_{-105}$ & $11.42^{+0.19}_{-0.32}$  & $-1.36^{+0.12}_{-0.12}$\\
\enddata
\end{deluxetable*}

The incorporation of eccentric effects produces notable shifts in some QC parameters, such as effective spin and component masses, due to their strong correlation with the length of the signal. As a consequence, the \phTE results with a uniform eccentricity prior (pink) favor lower mass values compared to the QC runs. This shift compensates for the shortening of the waveform induced by eccentricity, and consequently also impacts the inferred spin.
The use of a log-uniform prior suppresses the measured value of eccentricity due to its strong weight at low eccentricities. However, 
we observe that even a log-uniform prior in eccentricity yields a subdominant posterior mode at \mbox{$e_{20\mathrm{Hz}}\sim0.11$}. Although this support remains subdominant, it induces clear deviations from the QC results in the mass and spin distributions.
This observation is further supported by a nominally positive log-10 Bayes factor between the eccentric and QC hypotheses, \mbox{$\log_{10}\mathcal{B}_{\mathrm{E/QC}} = 0.11^{+0.11}_{-0.11}$}. 
While this value alone is not enough to claim strong evidence for eccentricity, it is still notable given the relatively low SNR of the event and the strong preference for low eccentricities imposed by the log-uniform prior.

\begin{figure}
    \centering
    \includegraphics[width=\linewidth]{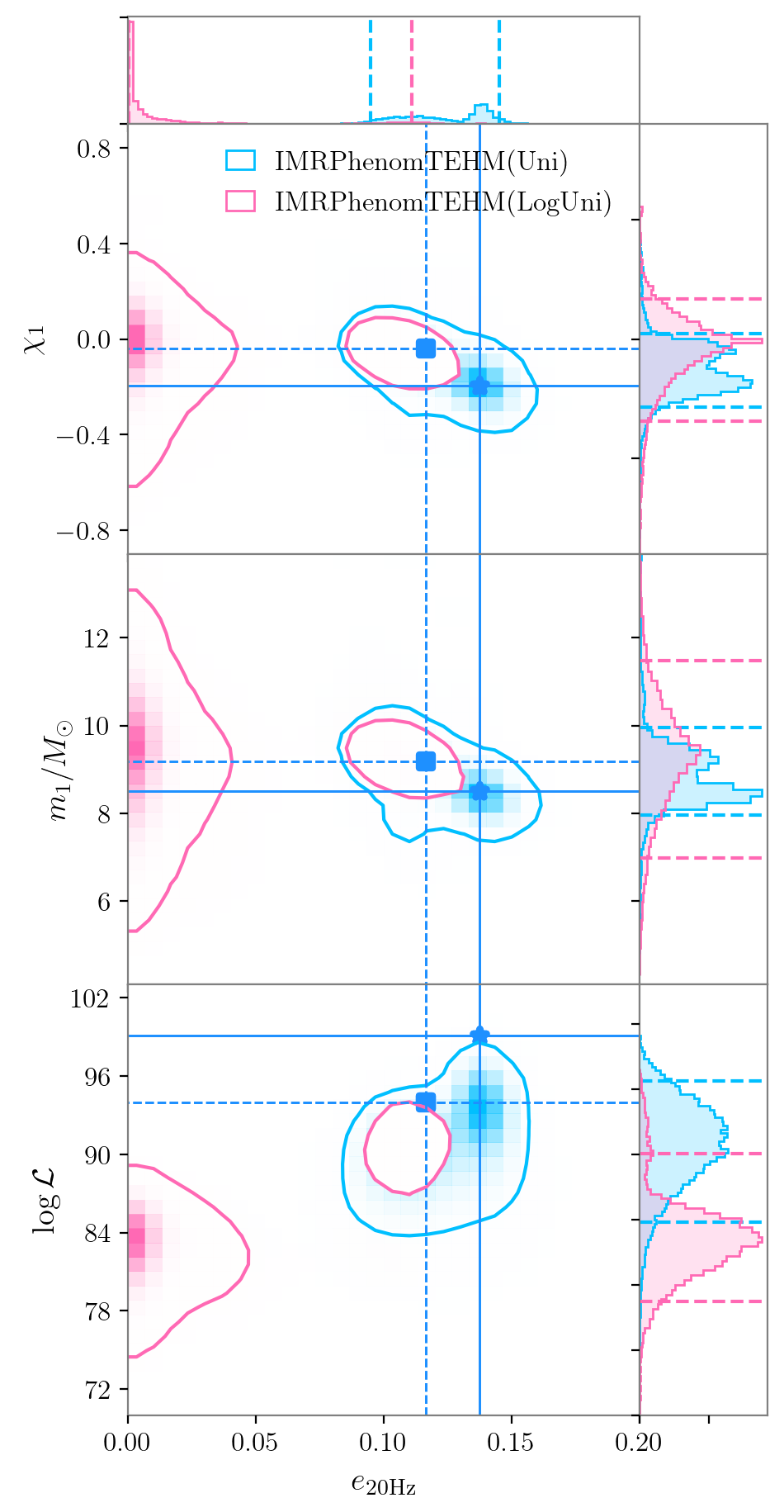}
    \caption{Marginalized one- and two-dimensional posterior distributions for GW200105, recovered using the \phTE model with two different eccentricity priors: uniform (Uni, green) and log-uniform (LogUni, blue). 
    The figure shows the joint distribution of the primary aligned spin component $\chi_1$ (\emph{top panel}), the primary mass $m_1/M_{\odot}$ (\emph{middle panel}), and the log-likelihood $\log \mathcal{L}$ (\emph{bottom panel}), as a function of the eccentricity measured at 20 Hz ($e_{20,\mathrm{Hz}}$).
    The blue star marks the maximum-likelihood sample from the uniform prior run, while the blue square indicates the second peak in the posterior distribution also present in the uniform prior run. 
    }
    \label{fig:degeneracies}
\end{figure}

Biases in the recovered QC parameters due to unmodeled eccentricity were already evident in earlier studies~\citep{Gupte:2024jfe, Planas:2025feq, Planas:2025jny, Morras:2025xfu}. 
For GW200105, our eccentricity measurement using a uniform prior reveals a bimodal posterior with peaks at \mbox{$e \sim 0.11$} and \mbox{$e \sim 0.14$}, corresponding to log-likelihood values of \mbox{$\log \mathcal{L} \sim 95$} and \mbox{$\log \mathcal{L} \sim 100$}, respectively. These modes are highlighted in Fig.~\ref{fig:degeneracies} by a star and a square.
This bimodality likely arises from correlations between eccentricity and QC parameters, such as the component masses and aligned spin, which influence the duration of the signal. Notably, the distinct peaks in the eccentricity posterior coincide with secondary peaks in the posteriors of the primary mass and the aligned spin component.
We find that the higher peak in eccentricity at \mbox{$e\sim0.14$} corresponds to a lower primary component mass \mbox{($m_1\sim8M_{\odot}$)}, which in turn implies a lower mass ratio of $q\sim0.25$ (see Fig.~\ref{fig:corner}). 
The reduced primary mass leads to a longer waveform, partially compensating for the signal shortening caused by eccentricity. Notably, this eccentric solution is also associated with a more strongly anti-aligned primary spin component \mbox{($\chi_1 \sim -0.2$)}, in contrast to the near-zero aligned spin obtained in the quasi-circular case. Since an anti-aligned spin tends to shorten the inspiral, this highlights a nontrivial interplay between eccentricity, mass, and spin in shaping the waveform duration and its frequency evolution.

Besides correlations and degeneracies between eccentricity, masses and spins, there are several explanations that can account for the observed bimodality in the eccentricity posterior.
Firstly, the relatively low network matched-filtered SNR of GW200105 ($\sim$14) inherently complicates the interpretation of this event and the determination of the dominant source of variation in the waveform frequency and duration. 
However, we note that the same data was analyzed in~\citep{Morras:2025xfu} without observing such a bimodality
and in the following, we report several tests performed to systematically assess the differences found. Across all variations we observe that the bimodality in the eccentricity posterior persists.

\begin{figure}
    \centering
    \includegraphics[width=\linewidth]{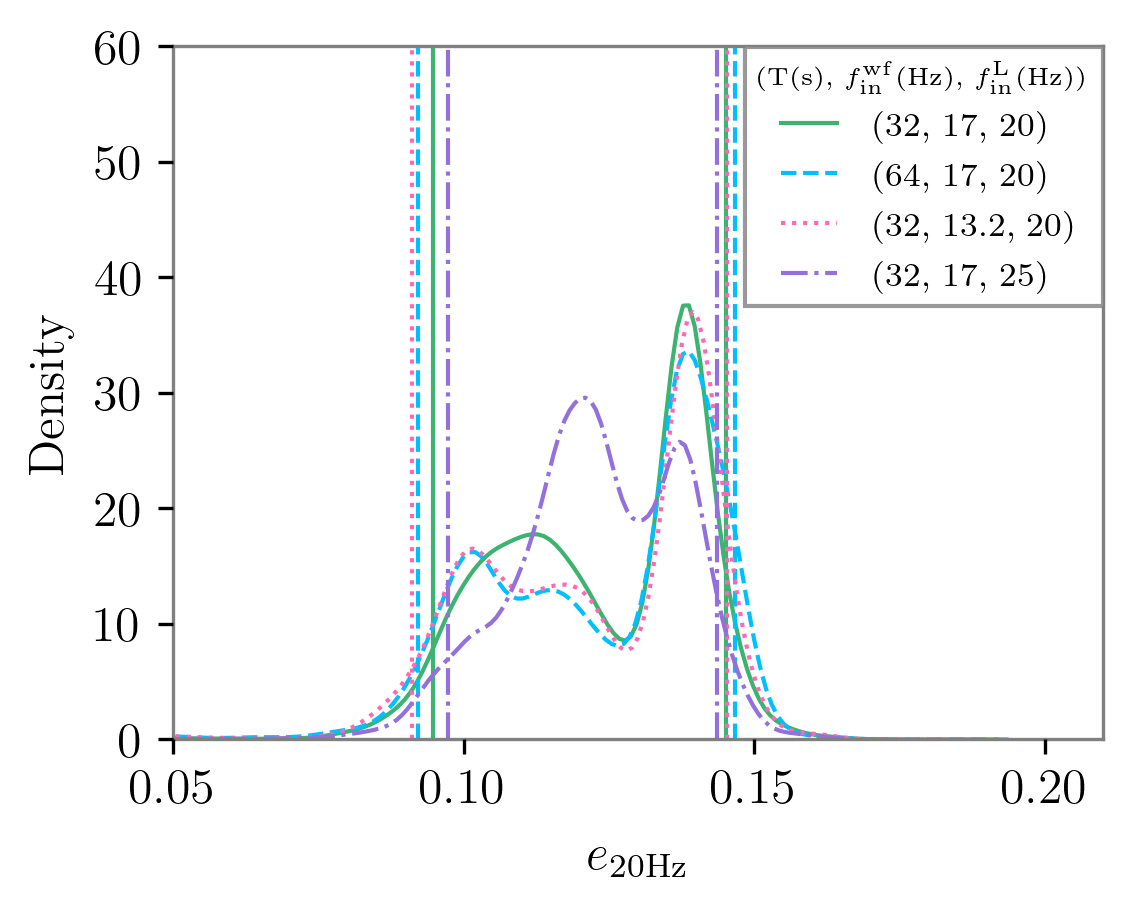}
    \caption{Posterior density distributions for the eccentricity at 20 Hz, $e_{\mathrm{20Hz}}$, using the \phTE (TEHM, or TE when restricting to the $(2,2)$-mode) model under different prior choices and starting frequencies. Solid lines correspond to a starting frequency of \mbox{$17\,\rm{Hz}$} (default runs, see Tab.~\ref{tab:GW_pes}), dashed lines to \mbox{$20\,\rm{Hz}$}, dotted to \mbox{$13.3\,\rm{Hz}$}, and dot-dashed lines.}
    \label{fig:eccvalues}
\end{figure}

Investigating whether spin-precession could be responsible for this behavior is challenging.
The posterior reported in~\citep{Morras:2025xfu} for the effective precession parameter, \mbox{$\chi_{\mathrm{p}} = 0.06^{+0.13}_{-0.05}$}, indicates only weak support for precession, making it unlikely to significantly alter the results. Notably, the inferred eccentricity remains very similar across both precessing and non-precessing analyses.
To better understand the nature of the two peaks observed in our eccentricity posterior, we computed mismatches between the corresponding waveforms shown in Fig.~\ref{fig:degeneracies}, using the PSDs from the detection. While the mismatch between the unprojected waveforms is relatively large (\mbox{$\mathcal{M} \sim 0.4$}), it reduces significantly when the waveforms are projected onto the detector response, yielding \mbox{$\mathcal{M} \sim 0.06$}.
Since this low mismatch is model-dependent, it is plausible that introducing an additional degree of freedom, such as spin precession, could help break the degeneracy, as seen in~\citep{Morras:2025xfu}.
We further analyzed the accumulation of SNR and log-likelihood as a function of frequency for both peaks to assess whether the inspiral-only setup in~\citep{Morras:2025xfu} could break the degeneracy. However, we found that most of the contribution comes from the $30$--$400\,\mathrm{Hz}$ range -- indicating that the merger-ringdown region does not significantly influence the results, and that the early inspiral, where the two waveforms are more different, is not the primary driver of signal recovery.
To assess the impact of known low-frequency noise contamination in LIGO Livingston, we performed an additional run in which the likelihood integration for LIGO Livingston started at 25 Hz. This run still yielded a bimodal posterior with uncertainties consistent with the original analysis as seen in Fig.~\ref{fig:eccvalues}, reinforcing the robustness of the previous observations.
As an additional test, we performed a zero-noise injection using the maximum likelihood parameters from the default \phTE run with a uniform prior in eccentricity, employing the Advanced LIGO and Virgo PSD curves. The recovered eccentricity posterior does not exhibit any bimodality, suggesting that the observation in the real data may be driven by noise or model degeneracies, consistent again with the findings above.

Finally, we also investigate possible systematics due to the signal duration and low frequency resolution, further developed in Appendix~\ref{ap:FDecc}.
To test the potential impact of harmonic content at 20 Hz, we compare the difference between starting the waveform generation at \mbox{$17\,\rm{Hz}$}, 
and \mbox{$13.3\,\rm{Hz}$} to ensure that all $l \leq 3$ modes are in band by \mbox{$20\,\rm{Hz}$}.
As shown in Fig.~\ref{fig:eccvalues}, the bimodality in the eccentricity posterior persisted in both cases. 

Finally, we find no significant improvement in convergence when lowering the starting frequency, suggesting that the limited duration constrains sensitivity to low-frequency content. The LVK assigns a 32-second duration to GW200105, which limits frequency resolution below \mbox{$30,\rm{Hz}$}. An extended 64-second run failed to resolve frequencies below \mbox{$25,\rm{Hz}$} and yielded a posterior consistent with the previously observed bimodality.
Further extending the duration is not feasible, as the cleaned Virgo data used in our analysis is limited to 64 seconds. Given that prior tests also suggest low-frequency content has minimal impact on the observed features, we do not expect this limitation to significantly affect our results. Nonetheless, it highlights an important consideration for future observing runs using eccentric models.

\section{Conclusions}\label{sec:conclusions}

This work presents the first PE analysis of NSBH events using a full IMR eccentric waveform model, marking a significant step forward in the complete interpretation of GW events. Our analysis with \phTE finds that GW200115 and GW230529 are consistent with QC NSBH inspirals, while GW200105 shows compelling evidence for non-zero eccentricity. This supports recent claims made in \citep{Morras:2025xfu} using an inspiral-only, eccentric, spin-precessing waveform model. While some systematic uncertainties remain, our results provide evidence in favor of eccentricity and signal the beginning of a new era in which robust eccentricity measurements are becoming feasible. To complement our results, we provide a data release~\citep{zenodo_release_NSBH} containing all the posterior samples generated in this work.

The outcome of all our investigations of GW200105 indicates that the \phTE model supports a non-zero eccentricity, with an estimated value of \mbox{$e_{20\mathrm{Hz}}=0.12^{+0.02}_{-0.03}$} and a Bayes factor of 
\mbox{$\log_{10}\mathcal{B}_{\mathrm{E/QC}}=1.22^{+0.12}_{-0.12}$} under a uniform prior. When using a more restrictive log-uniform prior, we still find a subdominant mode at non-zero eccentricity, also modestly supported with \mbox{$\log_{10}\mathcal{B}_{\mathrm{E/QC}}=0.11^{+0.11}_{-0.11}$}.
Although the log-uniform prior may not be well suited for low-SNR events such as this, the presence of a subdominant mode consistent with eccentricity is expected and notably was only observed in \citep{Morras:2025xfu} when using a higher lower bound for the eccentricity prior.

Runs with a uniform eccentricity prior show a bimodal posterior. Nonetheless, all tested configurations consistently support non-zero eccentricity with similar confidence intervals. Our tests indicate that the bimodality arises from a degeneracy between waveform projections onto the detectors, which could be broken by introducing additional parameters -- likely explaining why it vanishes when using the precessing eccentric model \texttt{pyEFPE}. Our investigations also highlight the importance of longer waveform durations to resolve low-frequency content for eccentric analysis, reinforcing the need for extended-duration analyses in future studies. A detailed summary of relevant systematics is provided in Appendix~\ref{ap:FDecc}.

Finally, it is worth highlighting that data quality may also influence the analysis of GW200105, as the event was affected in both detectors. 
This noise was subtracted from the publicly available strain data using the BayesWave algorithm~\citep{Cornish:2020dwh}, and we have confirmed that it does not affect the eccentricity findings.
Notably, the only other event in GWTC-3 reanalyzed with IMR eccentric models that shows strong evidence for eccentricity -- GW200129~\citep{gwtc3, Gupte:2024jfe, Planas:2025jny} -- is also known to have data quality concerns~\citep{Payne:2022spz}. While both events retain support for eccentricity even after applying various glitch mitigation strategies or excluding the affected data segments, these issues underscore the importance of carefully assessing data quality in future analyses.
Nonetheless, the results presented here provide valuable insights into the population properties and formation channels of compact binary mergers, marking an important step forward in GW astrophysics.

\begin{acknowledgments}
The authors sincerely thank Cecilio García-Quirós, Héctor Estellés, and Maria Haney for valuable discussions; and Alicia M. Sintes for her insights on frequency-domain resolution for long signals.
We also thank Geraint Pratten for his very helpful review of the manuscript as part of the LSC Publication \& Presentation Committee.
We thankfully acknowledge the computer resources (MN5 Supercomputer), technical expertise and assistance provided by Barcelona Supercomputing Center (BSC)  through funding from the Red Española de Supercomputación (RES) (AECT-2024-3-0027); and the computer resources (Picasso Supercomputer), technical expertise and assistance provided by the SCBI (Supercomputing and Bioinformatics) center of the University of Málaga (AECT-2025-1-0035, AECT-2025-1-0034).
This research has made use of data or software obtained from the Gravitational Wave Open Science Center (gwosc.org), a service of the LIGO Scientific Collaboration, the Virgo Collaboration, and KAGRA.
This material is based upon work supported by NSF's LIGO Laboratory which is a major facility fully funded by the National Science Foundation.
LIGO is funded by the U.S. National Science Foundation. Virgo is funded by the French Centre National de Recherche Scientifique (CNRS), the Italian Istituto Nazionale della Fisica Nucleare (INFN) and the Dutch Nikhef, with contributions by Polish and Hungarian institutes.

Maria de Lluc Planas is supported by the Spanish Ministry of Universities via an FPU doctoral grant (FPU20/05577, EST24/00621).
A. Ramos-Buades is supported by the Veni research programme which is (partly) financed by the Dutch Research Council (NWO) under the grant VI.Veni.222.396; acknowledges support from the Spanish Agencia Estatal de Investigación grant PID2022-138626NB-I00 funded by MICIU/AEI/10.13039/501100011033 and the ERDF/EU; is supported by the Spanish Ministerio de Ciencia, Innovación y Universidades (Beatriz Galindo, BG23/00056) and co-financed by UIB.
Jorge Valencia is supported by the Spanish Ministry of Universities via an FPU doctoral grant (FPU22/02211).
This work was supported by the Universitat de les Illes Balears (UIB); the Spanish Agencia Estatal de Investigación grants PID2022-138626NB-I00, PID2019-106416GB-I00, RED2022-134204-E, RED2022-134411-T, funded by MCIN/AEI/10.13039/501100011033; the MCIN with funding from the European Union NextGenerationEU/PRTR (PRTR-C17.I1); Comunitat Autonòma de les Illes Balears through the Direcció General de Recerca, Innovació I Transformació Digital with funds from the Tourist Stay Tax Law (PDR2020/11 - ITS2017-006), the Conselleria d’Economia, Hisenda i Innovació grant numbers SINCO2022/18146 and SINCO2022/6719, co-financed by the European Union and FEDER Operational Program 2021-2027 of the Balearic Islands; the “ERDF A way of making Europe”.
\end{acknowledgments}

\begin{contribution}
Maria de Lluc Planas is the primary developer of the \phTE model, coordinated and distributed the PE runs, and played a central role in shaping the overall narrative of the work. She led the writing of the manuscript, generated the graphical content, and managed the submission process. 

Sascha Husa has taken the initiative to develop this work, performed PE runs, and contributed to writing the manuscript.

Antoni Ramos-Buades is one of the main developers of the \phTE model, and contributed to writing the manuscript.

Jorge Valencia contributed to the software setup of \texttt{parallel\_bilby} on the MareNostrum5 supercomputer and conducted PE runs on the same system. He also reviewed and edited the manuscript.


\end{contribution}

%

\software{
    \texttt{phenomxpy}~\citep{Garcia-Quiros:2020qpx},
    \texttt{numpy}~\citep{numpy},
    \texttt{numba}~\citep{numba},
    \texttt{scipy}~\citep{scipy},
    \texttt{matplotlib}~\citep{matplotlib},
    \texttt{corner}~\citep{corner},
    \texttt{bilby}~\citep{bilby},
    \texttt{bibly\_pipe}~\citep{bilby_pipe_paper},
    \texttt{parallel\_bilby}~\citep{parallel_bilby_paper},
    \texttt{dynesty}~\citep{dynesty},
    \texttt{gwosc}~\citep{gwosc12,gwosc3}.
}


\appendix

\section{Waveform systematics in eccentric gravitational wave parameter estimation}\label{ap:FDecc}

Likelihood-based PE analyses in GW astronomy are typically performed in the frequency domain for ground based detectors, where events are sufficiently short that the response and noise power spectral density can be approximated as constant during an event, and the likelihood integral is then performed in the frequency domain. For time-domain waveform models, this requires special considerations, as a fixed starting time translates to different starting frequencies for different harmonics. Ensuring that all relevant harmonics are within the detector band at the beginning of the likelihood integration demands that the waveform starts sufficiently early in time. While this requirement applies to any multipolar time-domain model, it is especially important for eccentric systems, where each multipole contains a superposition of harmonics of the mean anomaly. Accurately capturing these components can significantly increase computational cost, particularly for low-mass binaries.
In contrast, purely frequency-domain models do not share these constraints, making them computationally cheaper to ensure that the information content of all multipoles is used consistently. 

For generic time-domain multipolar models, incorporating higher-order modes requires lowering the starting frequency to ensure that all relevant modes are excited at the beginning of the analysis band, typically set to \mbox{$20\,\rm{Hz}$} for current ground-based detectors. Since modes with \mbox{$m > 2$} contain higher-frequency content over the same time interval, the waveform must start earlier to capture them. The appropriate starting frequency can be adjusted based on the highest \mbox{$m$}-mode (\mbox{$m_{\max}$}) included in the analysis, following
\begin{equation}
f_{\mathrm{min}}^{\mathrm{wf}}=\frac{2f_{\mathrm{start}}^{\mathrm{L}}}{m_{\max}}.
    \label{eq:fstart}
\end{equation}
Hence, to ensure that multipoles with \mbox{$l \leq 3$} (as considered in this study) are within band at the likelihood starting frequency of \mbox{$20\,\rm{Hz}$}, waveform generation would ideally begin at \mbox{$13.3\,\rm{Hz}$}. However, after performing several tests and evaluating the impact on GW200105 (see Tab.~\ref{tab:GW_pes} and Fig.~\ref{fig:eccvalues}), we found that including such low frequencies had a negligible effect compared to the increased computational cost. Therefore, we set the default starting frequency for waveform generation to \mbox{$17\,\rm{Hz}$} in all runs.

Eccentric waveforms can be decomposed in harmonics of the mean anomaly, which describe the morphology of the signal~\citep{PhysRev.136.B1224}. Using the Stationary Phase Approximation (SPA)~\citep{spa} one can estimate the starting frequency where the $j$-th mean anomaly harmonic, $j\in (-N,+N)$ with $N\in \mathbb{N}$, enters in band,
\begin{equation}
    f_j \approx  \left(1 + \frac{j}{2} \right)f_{\mathrm{start}}^{\mathrm{wf}}.
    \label{eq:fstart2}
\end{equation}
The approximation in Eq.~\eqref{eq:fstart2} becomes more inaccurate for $j<-2$ where the $f_j = 0$ (numerical evolutions show that $f_j < 1$). 
The conclusion of Eq.~\eqref{eq:fstart2} is that including the full content of the positive harmonics at 20 Hz would require lower frequencies. However, for the parameter space considered and some waveform inspections, the harmonics with $j>2$ show a negligible contribution.

A key difference between \texttt{pyEFPE} and \phTE is that the former uses the SPA to derive closed-form expressions for the waveform modes, directly evaluated in the frequency domain. In contrast, time-domain models such as \phTE require a numerical transformation to the frequency domain. This demands additional care to ensure that all relevant modes are excited within the analysis band, as discussed above.
However, our tests confirm that setting a sufficiently low starting frequency for waveform generation in \phTE yields a clean Fourier transform and results in frequency-domain waveforms consistent with those from \texttt{pyEFPE}. Thus, differences in our PE results cannot be attributed to this modeling choice.
Instead, these investigations revealed a more critical issue for eccentric PE: the limited signal duration assigned to GW200105 (32 seconds) restricts the frequency resolution in \texttt{bilby}, where $\mathrm{d}f = 1/D$, making it challenging to capture low-frequency content below approximately 30 Hz. 
We found that only a significantly longer duration -- around $D\sim 128$ seconds -- produced a clean and well-resolved frequency-domain waveform.
While this low-frequency range has limited impact on the overall SNR or likelihood accumulation for this event, we still observe variations in posterior structure when modifying the waveform content between \mbox{$20–30\,\rm{Hz}$}, as shown in Fig.~\ref{fig:eccvalues}. This suggests that even when low-frequency contributions do not dominate signal reconstruction, they can still impact parameter inference.
This is particularly relevant for eccentric binaries, which contain more low-frequency structure than QC systems, since eccentricity is higher at low frequencies. While current detectors and low-SNR events like GW200105 may not benefit substantially from these low frequencies, future high-SNR detections and more sensitive detectors may require longer waveform durations to fully capture low-frequency content. This emphasized the need for extended-duration analyses in future studies where eccentric waveform models might be routinely used, and more towards next observatories more sensitive in the low frequencies.

\bibliography{sample7}{}
\bibliographystyle{aasjournalv7}

\end{document}